\documentclass[10pt,twocolumn]{article}
\usepackage[utf8]{inputenc}
\usepackage[T1]{fontenc}
\usepackage[margin=0.85in]{geometry}
\usepackage{times}
\usepackage{amsmath,amssymb}
\usepackage{graphicx}
\usepackage{booktabs}
\usepackage{multirow}
\usepackage{xcolor}
\usepackage[numbers,sort&compress]{natbib}
\usepackage[hidelinks]{hyperref}
\usepackage{caption}
\usepackage{pifont}
\graphicspath{{figures/}}
\title{\vspace{-1.0em}\bfseries Distribution-First Population Simulation:\\
Collapse, Calibration, and Recall in Non-WEIRD LLM Persona Modeling}
\author{G\"urkan \"Ozkan\\ Istanbul Technical University\\ \texttt{ozkangu@itu.edu.tr}}
\date{}

\begin{document}
\maketitle

\begin{abstract}
Synthetic-population tools increasingly run every individual as an independent large language model (LLM)
agent. Using real survey microdata, we show that this paradigm has a basic failure mode, and we set a
distribution-first corrective against it. Everything is measured with a deterministic, construct-validated
verifier on non-WEIRD (Turkey-first) data. First, $N$ independent LLM agents grounded on 2{,}414 real World
Values Survey respondents fail to reproduce the population's response \emph{distribution}. They pile onto a
modal default: across four scenarios and five seeds, concentration rises $0.36\!\to\!0.69$, entropy falls
$1.46\!\to\!0.77$, $85\%$ of units collapse, and $\mathrm{TVD}{=}0.44$. The collapse is a predictable
function of scenario structure ($r{=}0.55$ with having a single normatively correct answer). Second,
Verbalized Sampling (VS) fixes the field's chronic under-dispersion without training, and it does so in three
separate model families (fidelity $+7$ to $+10$; significant on Qwen, paired over seeds, $p{=}0.002$, $d{=}6.2$). The same move, however, universally
overshoots into over-dispersion (SD-ratio $0.4\text{--}0.56\to 1.26\text{--}1.37$), which is a structural
property of VS rather than one model's quirk. Third, survey fidelity transfers only weakly to agentic
behavior. In a single-model, single-domain booking task, a survey-calibrated persona is dominated by a cheapest-default
(${\sim}80\%$) that persona income modulates but does not override (comfort choice rises monotonically
$0\%\to7\%\to32\%$ across low/mid/high income bands). Along the way we correct a metric that confounded
flight duration with price and, left uncorrected, had made two of three models look ``context-blind.'' Fourth,
a placebo-controlled memorization attack and an anonymized national-election backtest show that VS keeps its aggregate strength
while subgroup and individual claims are contaminated by recall and underdetermination. We close with the
corrective: model the distribution once (VS) and assign it to characters consistently at $O(1)$ cost, with a
budget-aware router whose honest AUC we report ($0.805$, not the tautological $1.0$ of a code-derived oracle).
The central contribution needs no realism claim, because it measures the \emph{internal} inconsistency of the
independent-agent route and the \emph{conditions} under which the distribution-first route calibrates.
\end{abstract}

\section{Introduction}
Simulating human behavior with LLMs has moved quickly from a provocation to an engineering practice since
Argyle and colleagues framed ``silicon sampling'' \citep{Argyle2023}. Two lines have since separated. One
conditions a persona and predicts a survey distribution \citep{Argyle2023,OpinionQA,SimBench}. The other
runs each individual as an independent agent in an observe--memory--reflect--plan--act loop
\citep{Park2023,Concordia,AgentSociety}, and it is this second line that underwrites the commercial promise
of ``synthetic populations,'' the idea that one can simulate thousands of realistic people and read off their
reactions.

\begin{figure}[t]
  \centering
  \includegraphics[width=\columnwidth]{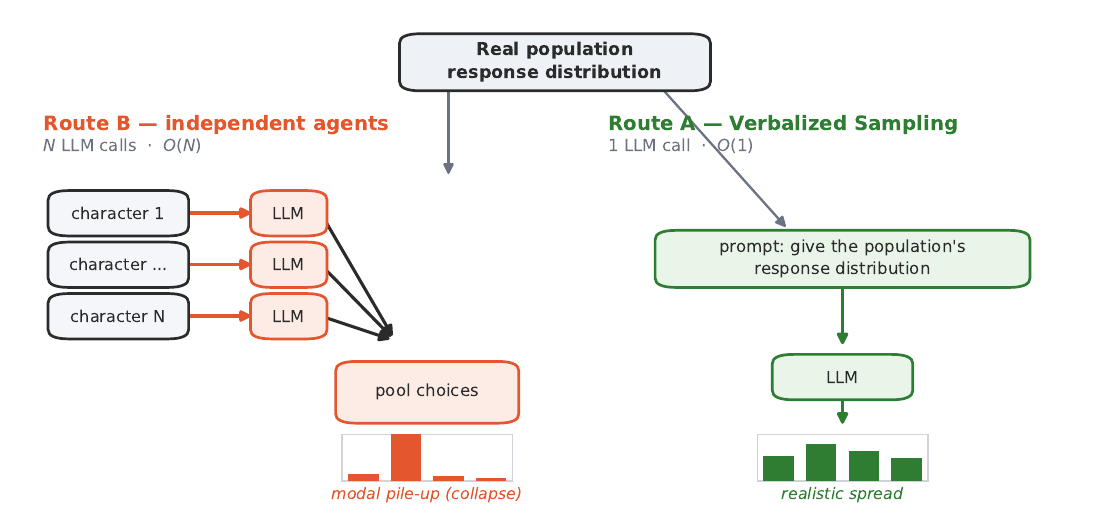}
  \caption{Two routes to a population distribution. Route~B runs each grounded character as an independent
  agent ($O(N)$ calls) and pools the choices, which pile onto a single option; Route~A (Verbalized Sampling)
  asks once for the whole distribution ($O(1)$) and keeps a realistic spread. Result~1 measures B's collapse
  relative to A.}
  \label{fig:routes}
\end{figure}

We ask a single question and answer it in four parts: does the independent-agent structure produce a
realistic population-level response \emph{distribution}, and if not, what does? Our answer to the first is no.
The failure is predictable rather than random, which is what makes it useful. Independent agents pile onto the
model's modal default instead of spreading across the population's real diversity, and the severity of that
pile-up depends on the structure of the scenario (Figure~\ref{fig:routes}). Our answer to the second is a
distribution-first recipe: model the distribution directly, then assign it to grounded characters.

Three commitments shape how we make these claims. First, we anchor every measurement to a deterministic
verifier whose own validity we test, so scores mean something and reproduce exactly. Second, we work non-WEIRD
(Turkey, real WVS microdata), because the field is overwhelmingly US/English \citep{Henrich2010} and because
Turkey adds a real access asymmetry worth documenting. Third, we run the primary pipeline on an open, local model and validate
cross-family generality on two others, a closed reasoning model (GLM-5.2) and a distinct open architecture
(Gemma-4-26B), so that ``no frontier needed'' does not rest on one model. The collapse result itself needs no
realism claim: we compare two \emph{internal} routes and measure one's collapse relative to the other, never
asserting that either simulates the real world. This keeps the central finding independent of the
memorization and data-access debates.

\section{Related Work}
\textbf{Silicon sampling and its limits.} \citet{Argyle2023} showed LLMs can produce demographically
conditioned ``silicon samples,'' and \citet{OpinionQA} and \citet{GlobalOpinionQA} mapped which groups models
reflect. The recurring finding is under-dispersion: the model flattens human diversity, a distributional counterpart
of the mode-seeking behavior long noted in open-ended generation \citep{Holtzman2020}. A mega-study found
twin-SD $<$ human-SD in 154 of 164 outputs, with individual correlation capped near $0.2$ \citep{MegaStudy},
and \citet{SimBench} pooled 20 datasets across 130+ countries and put the best model at ${\sim}40.8/100$, not
scaling with inference compute. A critical line documents deeper problems. Synthetic survey answers compress
variance, flip regression signs, and fail over time \citep{Bisbee2024}; $68$--$83\%$ of human-null effects
turn ``significant'' in synthetic data \citep{NatCompSci2025}; and of 53 studies, only one controlled for
leakage \citep{Anthis2025}. Identity prompting flattens groups \citep{Positionality}, and persona prompts
push toward caricature \citep{CoMPosT} and marked stereotypes \citep{MarkedPersonas}.

\textbf{Agent societies.} Generative Agents \citep{Park2023} introduced the observe--memory--reflect--
plan--act loop, Concordia \citep{Concordia} the Game-Master pattern, and AgentSociety \citep{AgentSociety}
$\sim$10k-agent experiments. These share the assumption that independent-agent interaction yields realistic
collective behavior, a premise that agent-based modeling has examined since well before LLMs
\citep{EpsteinAxtell1996,Bonabeau2002}. Recent LLM work converges on caution from other angles, notably that
role-conditioned agents suffer context collapse \citep{ContextCollapse}.

\textbf{Distribution-first and calibration.} The corrective we advocate has kin: Mixture-of-Personas
\citep{MixturePersonas} models a distribution and then assigns, and agentic economic modeling fits a bias
correction to a small human sample \citep{AgenticEcon}. Verbalized Sampling \citep{VS} is our training-free
lever. On agentic evaluation, $\tau$-bench \citep{TauBench} sets the tool-agent-user pattern, and simulated
users are known unreliable proxies \citep{LostInSim}.

\section{PersonaBench-TR and a Deterministic, Construct-Validated Verifier}
\label{sec:verifier}
\textbf{Grounded personas.} From the 2{,}414 real respondents of the WVS-7 Turkey wave \citep{WVS7} we
derive, for each, a deterministic character card: demographics, real value responses, computed features, and
a short first-person description. Generation costs nothing and inherits diversity from real heterogeneity
rather than from a high temperature. Because $77\%$ of the pool carries a unique value profile, generation
sidesteps the mode-collapse and ``Istanbul-educated default'' of LLM-generated pools \citep{MarkedPersonas}.

\textbf{The verifier.} \texttt{persona\_verify} scores deterministically (no seed): distributional distances
(TVD, KL, Jensen--Shannon, Wasserstein-1 \citep{PeyreCuturi2019}) and, kept separate, the SD-ratio
(model\_SD/human\_SD) that diagnoses under-dispersion. The headline \texttt{fidelity\_score} is $100\cdot(1{-}\mathrm{TVD})$; the
SD-ratio is never folded in, because burying variance in one number is the masking we criticize.

\begin{figure}[t]
  \centering
  \includegraphics[width=0.86\columnwidth]{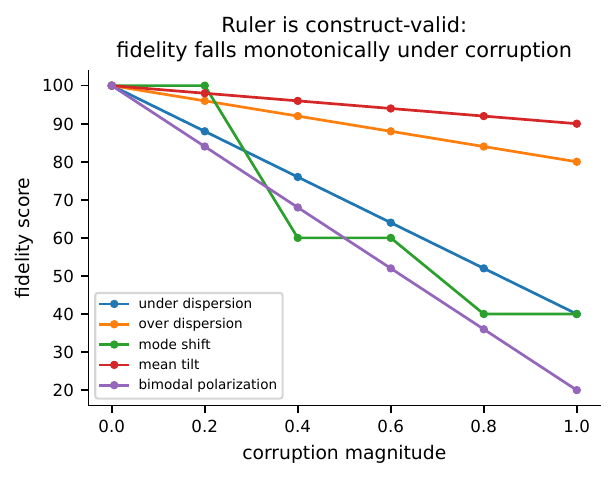}
  \caption{The verifier is construct-valid: fidelity falls monotonically as controlled corruption grows,
  across all five corruption families.}
  \label{fig:validity}
\end{figure}

\textbf{The ruler is itself validated.} Construct validity requires showing that a measure behaves as its
construct predicts \citep{CronbachMeehl1955}. With an LLM-free corruption suite we inject parametric distortions
into a known human distribution and check the ruler responds like a fidelity measure
(Figure~\ref{fig:validity}): fidelity falls monotonically across five corruption families; the SD-ratio is
direction-correct; Wasserstein is more ordinal-sensitive than TVD ($\Delta{=}1.60$ vs $0.60$); and a mildly
compressed distribution scores fidelity $88$ (looks excellent) while its SD-ratio of $0.89$ exposes the
variance collapse the single number hides. That last case is the empirical argument for reporting fidelity
and dispersion side by side.

\section{Result 1: Independent-Agent Populations Collapse, Predictably}
\label{sec:collapse}
Given a reaction scenario and $K$ discrete options, we compare two routes (Figure~\ref{fig:routes}).
\emph{Route A} (VS) asks the model directly for the population's probability distribution in one call,
averaged over three trials. \emph{Route B} (independent agents) has $N$ grounded characters each make one
choice, and the choices are pooled, for $N$ calls in total. Over four scenarios $\times$ five seeds ($=20$
units), $A{=}$VS and $B{=}$grounded(40), where each Route-B unit pools $40$ characters sampled from the
2{,}414-person pool. We summarize each pooled distribution by its \emph{concentration} (the modal probability
mass) and \emph{entropy} (Shannon entropy in nats), and call a unit \emph{collapsed} when Route-B
concentration exceeds Route-A's modal mass by more than $0.15$:

\begin{table}[h]
  \centering\small
  \begin{tabular}{lcc}
    \toprule
    Metric & Mean [95\% CI] & Reading \\
    \midrule
    TVD$(A,B)$ & $0.437$ [$0.383,0.498$] & B departs from A \\
    Concentration $A$ & $0.358$ [$0.345,0.374$] & realistic \\
    \textbf{Concentration $B$} & $\mathbf{0.685}$ [$0.614,0.764$] & pile-up \\
    Entropy $A$ & $1.464$ [$1.446,1.480$] & --- \\
    \textbf{Entropy $B$} & $\mathbf{0.770}$ [$0.614,0.906$] & diversity lost \\
    \textbf{Collapse rate} & $\mathbf{85\%}$ & \\
    \bottomrule
  \end{tabular}
  \caption{Independent agents pile onto a modal default (Figure~\ref{fig:collapse}).}
  \label{tab:collapse}
\end{table}

\begin{figure}[t]
  \centering
  \includegraphics[width=\columnwidth]{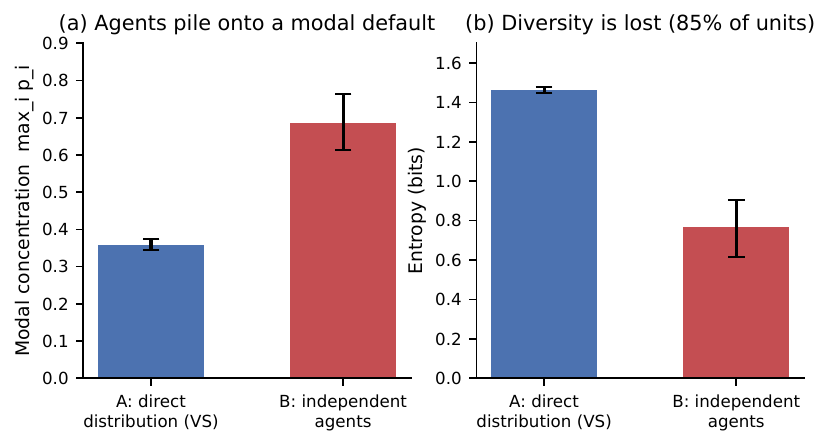}
  \caption{Route B (independent agents) concentrates and loses entropy relative to route A (direct VS
  distribution). Bars show bootstrap 95\% CIs over 20 units.}
  \label{fig:collapse}
\end{figure}

Independent agents, faced with the diverse distribution VS produces, gather on one option, usually the
``rationally correct'' one. This is the behavioral counterpart of survey under-dispersion \citep{MegaStudy},
and it generalizes across four scenarios rather than hinging on one. The collapse is not Qwen-specific. On
GLM-5.2 the same experiment gives $\mathrm{TVD}(A,B){=}0.418$ [$0.339,0.485$] ($n{=}12$) with the identical
structural pattern: single-rational scenarios collapse fully while the tradeoff scenario collapses least.

\begin{figure}[t]
  \centering
  \includegraphics[width=0.86\columnwidth]{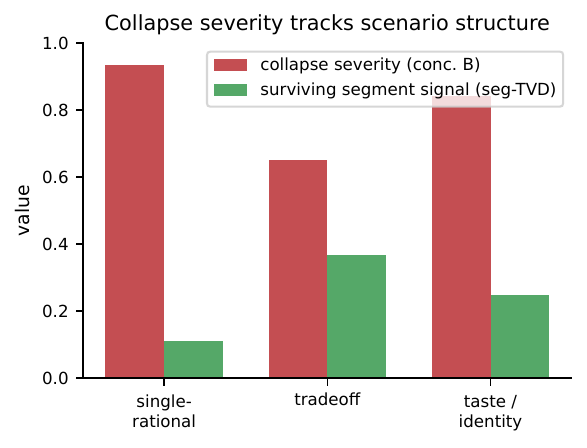}
  \caption{Collapse severity (modal concentration of route B) rises with single-rational structure, while the
  surviving between-segment signal rises with segment-relevance.}
  \label{fig:structure}
\end{figure}

\textbf{The collapse is structured, not uniform.} Labeling 16 scenarios into three structural classes
(\emph{single-rational}: a normatively correct answer; \emph{tradeoff}: value/budget dependent;
\emph{taste/identity}) over $16\times3$ seeds ($=48$ units, $25$ grounded agents per unit; Figure~\ref{fig:structure}) gives
three lessons. The right dependent variable is \emph{severity}, not a binary: the threshold is crossed almost
everywhere ($76$--$100\%$), so binary collapse correlates only weakly with structure ($r{=}0.26$), whereas
collapse \emph{severity} (concentration) correlates strongly with the single-rational structure
($r{=}0.55$). The surviving segment signal tracks segment-relevance (seg-TVD $0.11$ in single-rational,
$0.37$ in tradeoff, $r{=}0.45$), which is the direct rationale for the fix in \S\ref{sec:fix}. Unexpectedly,
taste/identity scenarios also collapse ($92\%$; food preference concentration $0.96$), so this is a general
modal-default tendency, most severe where a single answer is normatively privileged.

\section{Result 2: VS Calibrates Across Three Families, But Over-Disperses Universally}
\label{sec:vs}
\begin{figure*}[t]
  \centering
  \includegraphics[width=0.82\textwidth]{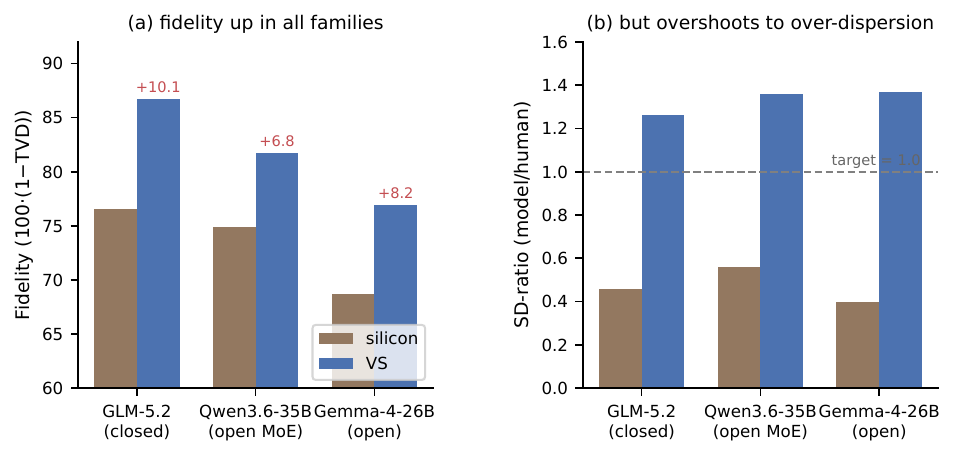}
  \caption{(a) VS lifts fidelity by $6.8$--$10.1$ points in three independent model families, so the method
  does not depend on the frontier. (b) The same move universally overshoots the target SD-ratio of $1.0$
  into over-dispersion, so VS is a structural over-corrector.}
  \label{fig:vs}
\end{figure*}

In the silicon baseline the SD-ratio sits near $0.5$. VS targets this directly: ask for the whole
distribution with probabilities, not one sample \citep{VS}. Whether it works on one model or a class decides the weight of the
claim, so we measured it in three families (Table~\ref{tab:vs}, Figure~\ref{fig:vs}).

\begin{table}[h]
  \centering\small
  \begin{tabular}{lcccc}
    \toprule
    Model & Sil.\ & VS & $\Delta$ & SD sil$\to$VS \\
    \midrule
    GLM-5.2 (closed)      & 76.6 & 86.7 & \textbf{+10.1} & $0.46\!\to\!1.26$ \\
    Qwen3.6-35B (MoE)     & 74.9 & 81.7 & +6.8 & $0.56\!\to\!1.36$ \\
    Gemma-4-26B (open)    & 68.7 & 76.9 & +8.2 & $0.40\!\to\!1.37$ \\
    \bottomrule
  \end{tabular}
  \caption{VS raises fidelity in all three families ($+6.8$ significant at $p{=}0.002$, $d{=}6.2$ on Qwen; paired over seeds)
  but throws every model from under- to over-dispersion.}
  \label{tab:vs}
\end{table}

Two things are universal here. VS raises fidelity by six to ten
points everywhere. But the same move throws the distribution from under-dispersion ($0.40$--$0.56$) to
over-dispersion ($1.26$--$1.37$): VS corrects variance collapse by pushing variance past the target, a
structural property rather than one model's quirk. VS should therefore be paired with a mean-preserving
correction. A plain sharpen pulls SD toward $1.0$ but costs four fidelity points (it distorts the mean),
whereas a KL-projection mean-preserving tilt pulls SD to $1.11$ while \emph{raising} fidelity $2.1$ points.
That the whole pattern repeats in three families makes ``over-dispersion is VS's signature'' consistent
across three independent families rather than a single model's artifact.

Two levers round out the design space. Backstory conditioning \citep{Anthology} moves \emph{backward}: a
$2\times2$ design shows a format-independent loss of $10$--$11$ points and SD collapsing to $0.20$--$0.37$,
as ``play this character'' shifts the model into role-play \citep{CoMPosT,Positionality}. Temperature does
nothing at all. Across $\{0.3\ldots1.3\}$ the SD-ratio stays at ${\sim}0.5$, because under-dispersion is a
\emph{between}-persona deficit while temperature is a \emph{within}-call loosening \citep{TempCreativity,
PriceOfFormat}.

\section{Result 3: Survey Fidelity Transfers Only Weakly to Agentic Behavior}
\label{sec:transfer}
Everything above is distribution-matching. The sharper question is whether fidelity survives a
\emph{decision}. We place the survey-calibrated persona in a tool-using booking task (IST$\to$BER) and log
decision-level fidelity step by step, following the $\tau$-bench persona-user pattern
\citep{TauBench}. A larger run ($n{=}60$, reasoning on, zero LLM-failures) resolves the picture by
income band (Table~\ref{tab:transfer}).

\begin{table}[h]
  \centering\small
  \begin{tabular}{lccc}
    \toprule
    income band & $n$ & ``cheapest'' & comfort \\
    \midrule
    low    & 2  & 1.00 & 0.00 \\
    mid    & 27 & 0.93 & 0.07 \\
    high   & 31 & 0.68 & \textbf{0.32} \\
    \midrule
    overall & 60 & 0.80 & 0.20 \\
    \bottomrule
  \end{tabular}
  \caption{GLM-5.2 persona-user booking, reasoning on ($n{=}60$). Behavior is cheapest-dominated, but persona
  income modulates it in the expected direction: comfort choice rises monotonically with income. Success
  $1.0$, budget-violation rate $0$, LLM-failure rate $0$. The low band is $n{=}2$, so its $0.00$ anchors the
  gradient weakly.}
  \label{tab:transfer}
\end{table}

The transfer is partial. Behavior is dominated by a cheapest-default
($80\%$ overall), so the model does \emph{not} simply reproduce the survey-calibrated persona. But the
default is not persona-blind either: comfort choice climbs monotonically from $0\%$ (low income) through
$7\%$ (mid) to $32\%$ (high). Survey-calibrated income thus reaches the agentic decision, but as a weak
modulation of a strong default rather than a faithful mapping. (An underpowered $n{=}30$ run had suggested a
near-null $7\%$ imprint; the larger run corrects this to a clear income gradient, an instance of the small-$n$
instability we flag throughout.) Tool-call validity is $100\%$ with no budget violation, so the constraint
lives in the world, not the agent's head. One methodological point the run surfaced is worth stating: a Qwen
reasoning-on booking with too small a token budget produced a $91\%$ LLM-failure rate, and because our
verifier separates LLM-failures from behavioral outcomes, this was flagged rather than silently scored as
``cheap.'' That is the kind of error-path hygiene that agentic transfer claims require. The finding
complements the fragility of simulated users \citep{LostInSim}.

\textbf{A metric trap, reported in full.} A metric meant to measure agentic context-sensitivity first led us
astray. It assumed a persona should buy the in-flight meal more often on a long flight, so
$\mathrm{meal(long)}-\mathrm{meal(short)}$ should be positive. GLM-5.2 returned zero, reading as ``even the
frontier model can't see context.'' But once we let the model give a one-sentence justification it behaved
exactly as expected, declining the meal on the short hop and adding it on the long haul. Two confounds had
combined. The design confounded duration with price: the ``long'' scenario was ten times more expensive
($16{,}000$ vs $1{,}600$~TL) and nearly exhausted the budget, so our modest-income personas rationally
declined extras, the opposite of the metric's assumption. The prompt confounded it further. An insistence
on ``give \textsc{this} person's decision, not your AI view'' bound the model so tightly to the frugal persona
that it suppressed the context signal, the agentic twin of the role-play absorption in \S\ref{sec:vs}.

\begin{figure}[t]
  \centering
  \includegraphics[width=0.86\columnwidth]{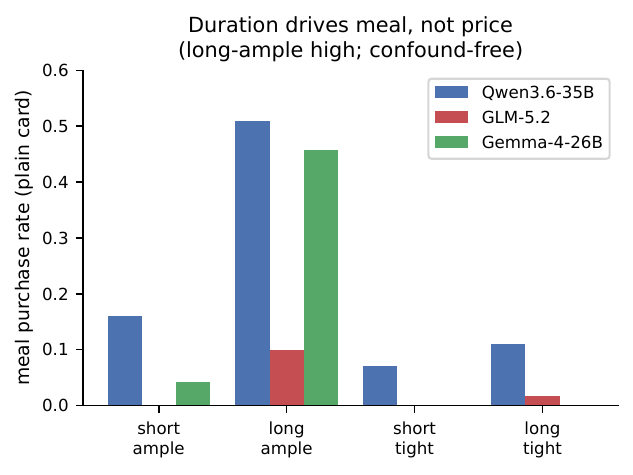}
  \caption{Meal-purchase rate by scenario (plain card). The long-ample bar is high while both tight-budget
  bars are near zero: duration drives the decision once the price confound is removed.}
  \label{fig:enrich}
\end{figure}

We isolated both by turning the design into a $2\times2$ (duration $\times$ budget-headroom), holding
affordability constant to read the \emph{pure duration effect}, with neutral framing
(Table~\ref{tab:enrich}, Figure~\ref{fig:enrich}).

\begin{table}[h]
  \centering\small
  \begin{tabular}{lccl}
    \toprule
    Model & meal(long$-$short) & entropy & enrichment \\
    \midrule
    Qwen3.6-35B ($n{=}100$) & \textbf{+0.35} & 1.96 & \textbf{harmful} \\
    Gemma-4-26B ($n{=}24$)  & \textbf{+0.42} & 1.20 & \textbf{harmful} \\
    GLM-5.2 ($n{=}60$)      & +0.10 & 1.02 & helpful \\
    \bottomrule
  \end{tabular}
  \caption{Corrected $2\times2$ context-sensitivity (pure duration effect, affordability held constant);
  enrichment here is an LLM-written backstory added to the grounded card.
  Both Qwen and Gemma read duration strongly; the old ``context-blind'' verdict was an artifact. The
  per-model direction (enrichment harmful for the strong-baseline pair, helpful for weak-baseline GLM) is
  stable across sample sizes ($n{=}24$ to $n{=}100$), though point estimates carry ${\sim}\pm0.1$ noise.}
  \label{tab:enrich}
\end{table}

The corrected picture inverts the first reading. Given the grounded card, both Qwen and Gemma read duration
context strongly (meal on the long flight rises from ${\sim}5$--$16\%$ to ${\sim}46$--$51\%$); the old
``context-blind ($0.0$)'' verdict was an artifact of the confound and role-play absorption, and it had made
two of three models (GLM and Qwen) look blind at once, even though Qwen's confound-free duration effect is in
fact strong. Cleared of the confound, the enrichment story rewrites itself:
LLM-written backstory helps only the model with a \emph{weak} baseline (GLM) and hurts the two with a strong
baseline (Qwen, Gemma), collapsing their decision diversity. In our three families, the value of enrichment
appears inversely related to baseline context-sensitivity: it acts as a \emph{substitute} that becomes
redundant, and damaging, once grounding is rich enough. This unifies with the survey side, where backstory already cost $10$--$11$ points, and
falsifies both ``enrichment always helps'' and ``enrichment always hurts.'' The broader lesson is procedural:
single-shot, JSON-only behavioral metrics can turn a real capability into an artifact.

\section{Result 4: Aggregate Fidelity May Be Recall}
\label{sec:memo}
The objection that VS is recall not reasoning deserves a direct answer, so we built a three-test,
placebo-controlled attack, run on Qwen (and cross-checked on GLM in the released artifacts) against real WVS-TR ground truth
(Table~\ref{tab:memo}).

\begin{table}[h]
  \centering\small
  \begin{tabular}{p{4.0cm}cc}
    \toprule
    Test (Qwen) & Result & Signal \\
    \midrule
    Conditional beats national recall? & $\Delta{=}{-}0.165$, $p{<}0.001$ & \ding{55} \\
    Subgroup gradient matches reality? & $62\%$, $\rho{=}0.68$ & weak \\
    Sensitive \emph{and} specific? & sens.\ 1/3; spec.\ 0/2 & \ding{55} \\
    \bottomrule
  \end{tabular}
  \caption{Memorization attack. The placebo control is decisive (below).}
  \label{tab:memo}
\end{table}

The placebo is the methodological heart. An earlier control counted counterfactual sensitivity, a response
to a \emph{relevant} manipulation, as evidence of reasoning. The placebo, an \emph{irrelevant} manipulation,
refutes it:
the real counterfactual gave $\Delta{=}{+}1.13$, the placebo $\Delta{=}{+}1.49$ (mean-opinion shift on the
item's 1--10 scale), i.e.\ \emph{larger}. So
``counterfactual sensitivity'' is non-specific prompt-reactivity; measuring sensitivity without a specificity
control is misleading. The verdict is honestly mixed: VS holds the national marginal (Result~2 stands), but
subgroup claims are explainable by recall and prompt noise \citep{SubgroupError,MegaStudy}.

\textbf{A real-event backtest} makes two failure modes concrete. Holding party preference out, we predict
the vote distribution of a past national election in the study country from values. The VS distribution
matches the official result at $\mathrm{TVD}{=}0.051$, ahead of a
person's own self-report ($0.095$), but the outcome is a well-known pre-cutoff result, so this is plain recall, not simulation.
The grounded-individual route collapses instead (the largest party's share jumps from a true $42.6\%$ to a predicted
$88.0\%$, $\mathrm{TVD}{=}0.454$), in every model and thinking mode. Two failures: one remembers, the other
collapses; neither simulates. The diagnosis is instructive: the features that fix the vote (identity and
candidate effects) are absent from the survey, so supporters of two ideologically distinct parties are, in
value-space, twins. That is concrete evidence for the $r{\approx}0.2$ individual ceiling
\citep{MegaStudy,Salganik2020}.

\begin{figure}[t]
  \centering
  \includegraphics[width=0.86\columnwidth]{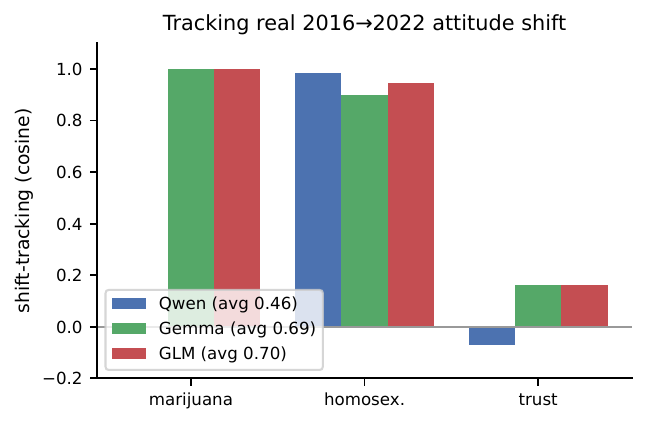}
  \caption{Tracking the real 2016$\to$2022 attitude shift. All three families track marijuana and
  homosexuality shifts well but share a trust-item blind spot.}
  \label{fig:temporal}
\end{figure}

\textbf{Temporal tracking across three families.} On three attitudes that genuinely shifted between the 2016
and 2022 GSS waves \citep{GSS}, we measure how well each model tracks the \emph{real} shift by cosine
(Figure~\ref{fig:temporal}): Qwen $0.46$, Gemma $0.69$, GLM $0.70$. The frontier and distinct open model
track shifts better, but all three share a blind spot: the trust item drifts toward ``it depends,'' a
class-level limit rather than one model's quirk. None of this replaces a true post-cutoff holdout; it
strengthens the case that such a holdout is an empirical necessity \citep{Anthis2025,Bisbee2024}.

\textbf{Grounding does generalize (not mere recall).} In a construct-holdout, the character is built from
five values and predicts ten held-out attitudes that never touched them. The value-aware card beats bare
demographics by $+3.6$ fidelity on GLM and $+2.9$ on Qwen ($10$ seeds, $p{=}0.002$, $d{=}2.24$). Because the
setup is BUILD$\neq$PREDICT, this is real generalization, a holdout-strengthened form of ``algorithmic
fidelity'' \citep{Argyle2023}. Grounding's contribution is real; its limit is at subgroup and individual resolution.

\section{Result 5: The Fix---Distribution-First With Character-Consistent Assignment}
\label{sec:fix}
\begin{figure}[t]
  \centering
  \includegraphics[width=\columnwidth]{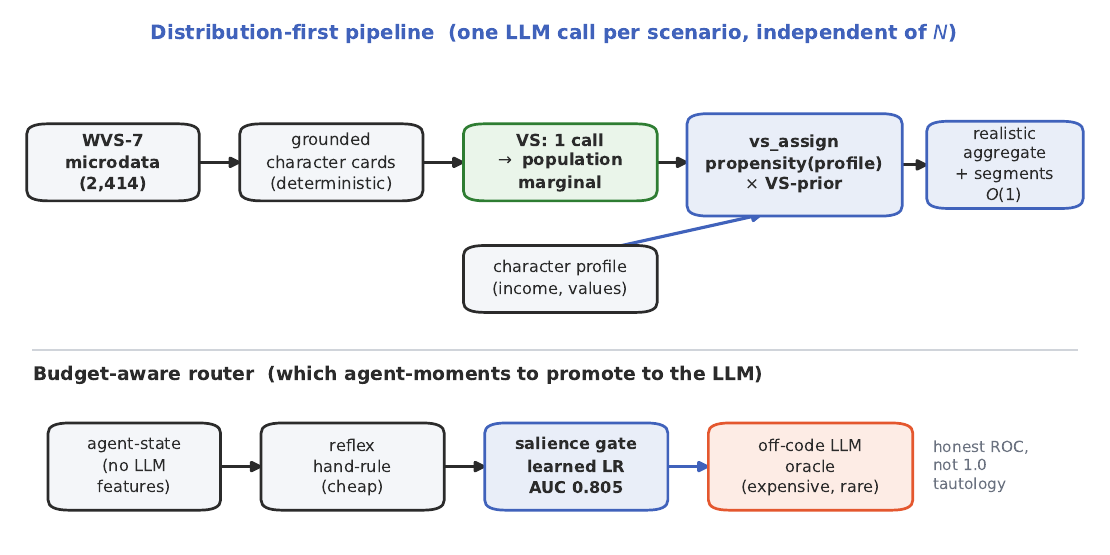}
  \caption{The distribution-first pipeline. VS produces the population marginal in one call; \texttt{vs\_assign}
  then gives each grounded character a response consistent with both the marginal and its own profile, at
  $O(1)$ cost. Below: the budget-aware router promotes only salient agent-moments to an off-code LLM oracle,
  with an honest AUC of $0.805$ rather than the tautological $1.0$ of a code-derived oracle.}
  \label{fig:pipeline}
\end{figure}

The structure results (\S\ref{sec:collapse}) point to the recipe (Figure~\ref{fig:pipeline}). Take the
population marginal with VS (realistic dispersion, no collapse), then \emph{assign} each character a response
consistent with both that marginal and their own profile, via propensity $\times$ VS-prior. This produces a
realistic aggregate with segment separation at \textbf{one LLM call per scenario}, independent of $N$. It
replaces the independent-agent route's $O(N)$ cost \emph{and} its collapse with $O(1)$ cost and a
distribution that does not collapse. It has kin in mixture-of-personas \citep{MixturePersonas}.

\textbf{A budget-aware router, measured honestly.} A hybrid architecture (cheap reflex + expensive LLM) must
decide which agent-moments to promote. The oracle must be an off-code LLM and the reflex a hand-rule, or the
separation is tautological. Our earlier evaluation drew the oracle from the reflex's own code and reported
$\mathrm{AUC}{=}1.0$, an artifact. With an off-code oracle, a learned logistic router (features computable
without an LLM at gate time, 5-fold CV) gives a real number: $0.805{\pm}0.074$ vs a fixed hand-rule $0.767$
on a local Qwen oracle. The learned weights overturn intuition: the most predictive features are
prior-entropy and reflex-uncertainty, not novelty or stakes. We do not overclaim. On a GLM oracle the fixed
rule beats the learned router ($\Delta{=}{-}0.17$), so the learned advantage does not generalize across
oracles. The durable win is methodological: the router AUC is now a real number, not a tautology.

\section{Discussion}
For the population-\emph{distribution} task, running every individual as an LLM is the wrong tool: independent
agents collapse to a modal default, predictably by scenario structure. The right tool models the distribution
directly (VS) and assigns it consistently, which is cheaper ($O(1)$) and collapse-resistant. Calibration needs
no frontier: the same VS gain and over-dispersion signature appear in three families. But survey fidelity does
carry into behavior only weakly, subgroup claims are contaminated by memorization, and individual prediction hits
underdetermination. Agent-society architectures \citep{Park2023,Concordia,AgentSociety} remain valuable for
narrative and interaction; for calibrated population-\emph{distribution} claims, the collapse must be
addressed explicitly, not assumed away.

\section{Limitations}
The collapse result characterizes route B; it does not prove route A (VS) is \emph{realistic}. That is the
job of external fidelity and temporal holdout, the one door that stays open (\S\ref{sec:memo} strengthens but
does not close it) \citep{Anthis2025,Bisbee2024}. VS over-disperses slightly and binary items resist full
calibration. We make no individual-twin claim ($r{\approx}0.2$ ceiling \citep{MegaStudy}). The router oracle
is itself an LLM and may carry memorization; its AUC measures internal fidelity/cost, not realism. Scale
beyond ${\sim}10$k agents and learned features are future work. The agentic-transfer finding covers one domain
(airline booking) and a single model (GLM-5.2) at adequate power. A Qwen reasoning-on replication was
underpowered (a tight token budget drove a high LLM-failure rate; \S\ref{sec:transfer}) and is left to future
work, and the metric confound was isolated on the meal dimension only. All
three families are evaluated in Turkish/English, so cross-lingual generality is untested.

\section{Ethics}
The paper's central messages are negative, and that is a safety property: we refute ``independent-agent
synthetic populations produce realistic distributions,'' document that survey fidelity transfers only weakly, and
show grounded-individual prediction collapses or recalls. Segment- and opinion-level outputs can be misused
for political or commercial manipulation; we make no individual-prediction claim and mark the limits
(backtest collapse/recall, temporal caveat) explicitly. Artifacts are released to measure \emph{what cannot be
simulated}; individual targeting and voter-profiling are out of license.

\section*{Reproducibility}
All scores come from a deterministic verifier; runs are seed-matched with a manifest. Code and results are
released with the character pool derived from WVS-7; a few exploratory runs are flagged \texttt{mock\_only}
in the release, and no headline number in this paper comes from a mock run. Full commands are in the
supplementary \texttt{REPRODUCE} guide.

{\small
\bibliographystyle{plainnat}
\bibliography{references}
}

\end{document}